\documentclass[aps,prb,twocolumn,showpacs,eqsecnum]{revtex4}
\usepackage{amssymb}
\usepackage{amsmath}
\usepackage{graphicx}

\begin{document}

\title{Relaxation in quantum systems. Manifestation of
the state-selective reactive decay}

\author{A. I. Shushin}
\affiliation{Institute of Chemical Physics, Russian Academy of
Sciences, 117977, GSP-1, Kosygin str. 4, Moscow, Russia\\
e-mail: shushin@chph.ras.ru}

\begin{abstract}
The effect of state-selective reactive decay on the relaxation
kinetics of quantum multistate systems is studied in detail in the
Bloch-Redfield approach (BRA). The results are applied to the
analysis of this effect in radical pair recombination kinetics. The
BRA is shown to be able to describe quantitatively most important
specific features of the recombination kinetics including those
predicted by phenomenological treatment and by recently proposed
approaches based on quantum measurement theories.
\end{abstract}

\maketitle

\bigskip

\section{Introduction}

Recently some attention has been attracted to the discussion of
theoretical approaches for describing the effect of state-selective
reaction on the reactive quantum systems. The most popular example
of such processes is spin-selective condensed phase physicochemical
reactions of paramagnetic particles.\cite{St} The interest in this
problem is inspired by new ideas, known in the quantum measurement
theory,\cite{Brag,Niel} which can probably be useful in the analysis
of relaxation in quantum systems.\cite{Hor,Kom} At first glance,
these ideas really provide some new insight into the specific
features of kinetics of state selective processes. In support of
this conclusion recently some results have been
presented,\cite{Gau,Hor,Kom} which seem to demonstrate some new
possibilities of the new methods for treating experimental results.

The discussion of these methods stems from some seeming limitations
of the traditional approach looking rather heuristic. The main point
of this approach is considered to be the description of
spin-selective reactivity.\cite{Hab} The description is based on the
simple model for the reactivity supermatrix $\hat K$ operating on
the spin density matrix of the system $\rho$. In this model, as
applied to radical pair (RP) recombination with the rates $\kappa_S$
and $\kappa_T$ in the singlet ($S$) and triplet ($T$) states,
respectively,
\begin{equation} \label{intr1}
\hat K \rho= \mbox{$\frac{1}{2}$}\big\{\kappa_{S}[P_S,\rho]_{+} +
\kappa_{T}[P_T,\rho]_{+} \big\}
\end{equation}
where $P_{S}$ and $P_T$ are the operators of projection on the $S$
and $T$ states, respectively, and $[P,\rho]_{+} = P\rho + \rho P$.
The analysis based on the quantum measurement approach shows
\cite{Hor,Kom} that the expression (\ref{intr1}) leads to the
experimentally distinguishable underestimation the dephasing effect
of reactivity.

In this short paper we will analyze the state-selective reactivity
effects on the relaxation in quantum systems within the
(traditional) Bloch-Redfield approach (BRA).\cite{Blo,Red,Hbr,Abr}
The BRA is the basic approach in the magnetic relaxation theory,
but, in principle, it is very general and under quite realistic
assumptions can properly describe a large variety of quantum
relaxation processes.

The results of our work demonstrate that formula of type of eq.
(\ref{intr1}) is directly follows from the BRA. It is, first,
derived for the simple three-state system (THSS), in which
relaxation in the two state system is induced by reactive
transitions to the third state, and then obtained [just in the form
of eq. (\ref{intr1})] as applied to the case of spin selective RP
recombination, by treating the reaction as spin-dependent
transitions from excited (electronic) $S$ and $T$ states of the RP
to the ground state.

It is also shown that the BRA enables one to essentially generalize
eq. (\ref{intr1}) by taking into account possible additional $ST$
dephasing, accompanying the reaction. The generalized formula
appears to be able to describe all predictions of the quantum
measurement approach, discussed in refs. [4,5], under some
assumptions on relative value of reaction and dephasing rates. The
effect of the interradical-distance dependent reactivity and
exchange interaction is also briefly analyzed within the approach
based on the stochastic Liouville equation \cite{Fre,Shu1} which
allows one to rigorously describe the manifestation of relative
diffusion of radicals.

\section{General formulas}

Here we briefly discuss the main points of the BRA which are
important for our further analysis of relaxation
kinetics.\cite{Hbr,Alek} For convenience, in our discussion we will
use the frequency units for energies (i.e. $\hbar = 1$) and the
notation $\beta = 1/(k_B T)$.

In the BRA the evolution of the system is described by the Liouville
equation for the density matrix $\rho (t)$:\cite{Fre,Shu1}
\begin{equation}
\dot \rho = -i[H_s,\rho] +\hat R \rho,\;\;\mbox{with}\;\;[H_s,\rho]
= H_s \rho - \rho H_s. \label{gen0}
\end{equation}
in which $H_s$ is the non-fluctuating part of the Hamiltonian $H$ of
the system and $\hat R$ is the relaxation supermatrix (the matrix in
the Liouville space of matrix elements of the density matrix). Thus,
the problem of description of the relaxation kinetics reduces to the
evaluation of the matrix $\hat R$ and subsequent solution of eq.
(\ref{gen0}).

For the system in a bath the Hamiltonian\cite{Hbr,Alek}
\begin{equation}
 H = H_s  + H_{sl} + H_{l} ,  \label{gen1}
\end{equation}
where $H_l$ is the Hamiltonian of the bath whose specific form is
not important for our further analysis. The term
\begin{equation}
H_{sl} =  \sum\nolimits_{n}\! \Phi_{n}({\bf q})\Lambda_{n}
\label{gen1b}
\end{equation}
represents the interaction of the system with the bath, in which
$\Lambda_n$ are the Hermitian operators in the space of the system
($|\nu\rangle ,\, \nu = 0, 1, 2$) and $\Phi_{n}(\bf{q})$  are the
lattice coordinate dependent Hermitian operators of fluctuating
amplitudes of interaction (see below).

In what follows it will be convenient to represent operators
$\Lambda_n$ as sums of those $\Lambda_n^r$, defined
by\cite{Hbr,Alek}
\begin{equation}\label{gen1c}
\Lambda_{n} (t) = e^{iH_s t}\Lambda_{n}e^{-iH_s t} =
\sum\nolimits_{r} \!\Lambda_{n}^{r} e^{i\omega_n^r t},
\end{equation}
which describe transitions between different pairs of states of
$H_s$ [$\Lambda_{n} \equiv \Lambda_{n} (t=0) = \sum\nolimits_{r}
\!\Lambda_{n}^{r}$].

In the sum (\ref{gen1c}) the frequencies $\omega_n^{r}$ are the
differences of eigenstates $\nu_j$ of the Hamiltonian $H_s$
corresponding to eigenstates $|j \rangle$ (i.e. $H_s |j \rangle =
\nu_j |j \rangle$):
\begin{equation}\label{gen1d}
\omega_n^{r} = \nu_j - \nu_{j'} = \omega_{jj'}
\end{equation}
and the operators $\Lambda_{n}^{r}$ are defined by
\begin{equation}\label{gen1e}
\Lambda_{n}^{r} = \langle j |\Lambda_{n}| j'\rangle
\delta_{\omega_n^{r},\omega_{jj'}}.
\end{equation}
The operators and  frequencies in eq. (\ref{gen1c}) satisfy the
relations $(\Lambda_{n}^{r})^{\dag} = \Lambda_{n}^{-r}$ and
$\omega_n^{-r} = - \omega_n^r$ (see examples below).\cite{Hbr,Alek}

In the BRA the relaxation rates  are determined by the pair
correlation functions\cite{Hbr,Alek}
\begin{equation}
K_{nn'}^{+} (t) = \langle \Phi_{n} (t)\Phi_{n'}\rangle, \; 
\; K_{nn'}^{-} (t) = \langle \Phi_{n} \Phi_{n'} (t)\rangle,
\label{gen2}
\end{equation}
in which
\begin{equation} \label{gen3}
\Phi_n (t) = e^{i H_{l}t}\Phi_n e^{-i H_{l}t}
\end{equation}
and the averages  $\langle \dots \rangle$ are evaluated over the
equilibrium lattice density matrix $\rho_{l}$:
\begin{equation} \label{gen4}
\langle A \rangle = {\rm Tr} (\rho_{l} A )
\;\;\mbox{with}\;\;\rho_{l} = e^{-\beta H_{l}}/{\rm Tr} (e^{-\beta
H_{l}}).
\end{equation}
In addition we will introduce their Fourier transforms
\begin{equation}\label{gen7}
J^{\pm}_{nn'} (\omega) = \int_{-\infty}^{\infty}\! dt \,
K_{nn'}^{\pm} (t)e^{i\omega t}
\end{equation}
for which we get $J_{nn'}^{+} (\omega) = e^{\beta \omega}
J_{n'n}^{-} (\omega) = e^{\beta \omega} J_{n'n}^{+} (-\omega).$

The most important BRA results are conveniently expressed in terms
of Fourier transforms
\begin{equation}
 J_{nn'} (\omega) = \mbox{$\frac{1}{2}$} [J_{nn'}^{+} (\omega) +
J_{n'n}^{-} (\omega)] \label{gen9}
\end{equation}
of symmetrized functions $K_{nn'} (t) = \mbox{$\frac{1}{2}$}
[K_{nn'}^{+} (t) + K_{n'n}^{-}(t)]$.

In particular, the supermatrix $\hat R$ is given by\cite{Hbr,Alek}
\begin{equation}\label{gen11}
\hat R  = \mbox{$\frac{1}{2}$}(\hat R_1 + \hat R_2),
\end{equation}
where
\begin{eqnarray}\label{gen11a}
\hat R_1 \rho \!&=&\!\!\!\sum_{nn'r}\! J_{n'n}(\omega_n^r)
\Big[\big[\Lambda_{n'}^r,\rho\big],\Lambda_n \Big], \\
\hat R_2 \rho \! &=&\!\! \!\sum_{nn'r}\!
J_{n'n}(\omega_n^r)\tanh(\mbox{$\frac{1}{2}$}\beta\omega_{n'}^r)
\Big[\Lambda_{n},\big[\Lambda_{n'}^r,\rho\big]_{+} \Big], \qquad\;
\end{eqnarray}
with $\big[\Lambda,\rho\big] = \Lambda\rho - \rho \Lambda
\;\;\mbox{and} \;\; \big[\Lambda,\rho\big]_{+} = \Lambda\rho + \rho
\Lambda.$

Strictly speaking, in addition to relaxation the BRA also predicts
some frequency shifts,\cite{Blo,Red,Hbr,Abr} though, we are not
going to analyze them in this work.

Concluding the short review of the BRA we recall the conditions of
validity of this approach. The most important is the condition
$\tau_c \ll \| \hat R\|$\cite{Blo,Red,Hbr,Abr} in which $\tau_c$ is
the characteristic correlation time of $H_{sl}$ fluctuations,
i.e.the average time of decay of correlation functions
$K_{nn'}^{\pm} (t)$. This condition insures the validity of the
second order perturbation approximation used in the derivation of
the kinetic equation (\ref{gen0}). Some additional condition is also
required which ensures fast thermalization of the lattice, and in
particular, fast relaxation of the non-diagonal diagonal elements
between $s$ and $l$ systems generated by the interaction
$H_{sl}$.\cite{Alek} This additional condition is, in reality, not
very restrictive\cite{Alek} and we are not going to discuss it in
this work.

\section{Three-state system}

In this Section we will demonstrate the effect of reactivity on the
quantum evolution kinetics by the example of the relaxation in the
simple THSS, schematically shown in Fig. 1a. The THSS consists of
the subsystem of two states $(|0\rangle, |1\rangle)$ and the third
state $|2\rangle$, transition to which model the reaction. In the
analysis we will also take into account the additional $0\!-\!1$
dephasing [because of possible fluctuations of $0\!-\!1$ splitting
(see Sec. III.C.)], but, first, we will neglect this effect.

\begin{figure}
\setlength{\unitlength}{1cm}
\includegraphics[height=5.5cm,width=8cm]{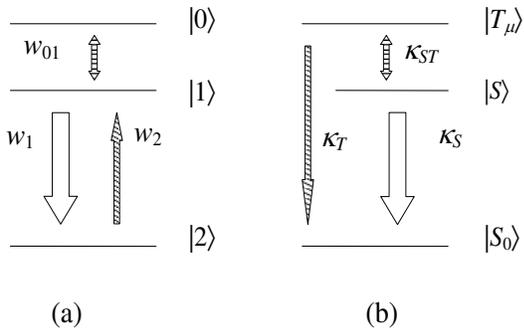}
\caption{The schematic picture of the levels  and transitions
between these levels for: (a) the three-state system ($w_1$ and
$w_2$ are the rates of population relaxation, and $w_{01}$ is the
$0\!-\!1$ dephasing rate) and (b) radical pair ($\kappa_S$ and
$\kappa_T$ are the reaction rates, and $\kappa_{ST}$ is the
$ST_{\mu}$ dephasing rate,\, ($\mu = 0,\pm$).}
\end{figure}

In the considered THSS the terms of the total Hamiltonian $H$ [eq.
(\ref{gen1})] are defined as follows
\begin{eqnarray}
H_s &=&   \omega_0 |0\rangle \langle 0| +
\mbox{$\frac{1}{2}$}\omega_s (|1\rangle \langle 1| - |2\rangle
\langle 2|), \label{tss1}\\
H_{sl} &=& H_{sl} (12) = \mbox{$\frac{1}{2}$} \!\sum\nolimits_{n =
x,y}\!\!\Phi_{n}({\bf q})\sigma_{n}. \label{tss2}
\end{eqnarray}

In the operator $H_{sl}$, representing (fluctuating) interaction
between $|1\rangle$ and $|2\rangle$ states, in which $\sigma_{n}, \,
(n = x,y),\,$ are the Pauli matrices in the space ($|1\rangle,
|2\rangle$): $\sigma_{x} = |1\rangle \langle 2| + |2\rangle \langle
1|$ and $\sigma_{y} = i(|2\rangle \langle 1| - |1\rangle \langle
2|)$ . In principle, $H_{sl}$ can be treated as the interaction of a
quasispin, associated with matrices $\sigma_{x,y}$, with the
fluctuating magnetic field, whose components are $\Phi_n ({\bf
q}),\, (n = x,y)$.

Formulas (\ref{tss1}) and (\ref{tss2}) present the most general form
of the two-state Hamiltonian in the absence fluctuating diagonal
interaction. As for the lattice coordinate dependent (fluctuating)
components  $\Phi_n ({\bf q})$, their particular forms are not of
real importance for our further analysis.

The operators $\Lambda_n^r$ introduced in formula (\ref{gen1c}) are
written as
\begin{eqnarray}\label{tss3}
\Lambda_x^1 \;\;&=& \;\:i \Lambda_y^{1} \;\;=\mbox{$\frac{1}{4}$}
(\sigma_x + i \sigma_y) =
\mbox{$\frac{1}{2}$}|1\rangle \langle 2|, \\
\Lambda_x^{-1} &=& \!\!-i \Lambda_y^{-1} =\mbox{$\frac{1}{4}$}
(\sigma_y - i \sigma_x) = \mbox{$\frac{1}{2}$}|2\rangle \langle 1| 
\end{eqnarray}

As for fluctuations of $\Phi_n$, for simplicity, they are suggested
to be uncorrelated and axially symmetric, i.e.
\begin{equation} \label{tss4}
J_{nn'} (\omega) = J(\omega) \delta_{nn'} , \;\; (n,n' = x,y).
\end{equation}

\subsection{Relaxation kinetics in two-state $(1,2)$ subsystem}

First of all, it is clear that in the absence of interaction between
the state $|0\rangle$ and two states $|1\rangle, |2\rangle$ the
relaxation in the subsystem of two states  $|1\rangle, |2\rangle$ is
described by the conventional Bloch equations for the density matrix
$\rho$ in the subspace $\{|1\rangle, |2\rangle\}$:\cite{Alek}
\begin{eqnarray}
\dot \rho_{11} &=&  - w_{11} \rho_{11} + w_{22} \rho_{22} \label{tss5a} \\
\dot \rho_{22} &=&  - w_{22} \rho_{22} + w_{11} \rho_{11}  \label{tss5b}\\
\dot \rho_{\mu} &=&  - w_{n} \rho_{\mu}, \qquad (\mu = 12,\, 21).
\label{tss5c}
\end{eqnarray}
The rates are given by
\begin{eqnarray} 
w_{11} &=& e^{\beta\omega_s}w_{22} =
J(\omega_s)/(1+e^{-\beta\omega_s}), \label{tss6}\\
w_n &=& w_{12} = \mbox{$\frac{1}{2}$}(w_{11} + w_{22}) =
\mbox{$\frac{1}{2}$}J(\omega_s). \label{tss7}
\end{eqnarray}

In particular, for large splitting $\omega_0 \gg \beta^{-1}$
\begin{equation} \label{tss8}
w_{11} = w_{11}^{\infty} = J(\omega_s), \;\;w_{22} = 0, \;\; w_{n} =
\mbox{$\frac{1}{2}$}w_{11}.
\end{equation}

\subsection{Relaxation kinetics in three-state system}

The analysis within the THSS shows that, in addition to relaxation
in $(|1\rangle, |2\rangle)$-subspace, the fluctuating interaction
$H_{sl}$ induces also the relaxation of $0\!-\!1$ and $0\!-\!2$
phases, i.e. the decay of non-diagonal elements $\rho_{0\mu}$ and
$\rho_{\mu0}$ with $\mu = 1,2$.

The expressions for corresponding phase relaxation can
straightforwardly be derived with the use of eqs. (\ref{gen11}) and
(\ref{gen11a}).

For example, the decay rate of $\rho_{01}$ is determined by the
matrix elements
\begin{equation} \label{tss9}
\langle 0|\hat R_1 \rho |1 \rangle = J(\omega_s)\langle
1|\Lambda_{xy}^{0}|1 \rangle \rho_{01} =
-\mbox{$\frac{1}{2}$}J(\omega_s) \rho_{01},
\end{equation}
and
\begin{eqnarray} \label{tss10}
\langle 0|\hat R_2 \rho |1 \rangle &=&
J(\omega_s)\tanh(\mbox{$\frac{1}{2}$}\beta\omega_s)\langle
1|\Lambda_{xy}^{0} |1 \rangle \rho_{01}\nonumber \\
& = & -\mbox{$\frac{1}{2}$}J(\omega_s)
\tanh(\mbox{$\frac{1}{2}$}\beta\omega_s) \rho_{01}
\end{eqnarray}
where
\begin{equation} \label{tss11}
\Lambda_{xy}^{0} =
\Lambda_x^{1}\Lambda_x^{-1}+\Lambda_y^{1}\Lambda_y^{-1}.
\end{equation}
Similar formulas can be obtained for $\langle 1|\hat R_{j} \rho |0
\rangle$ elements. Substitution of expressions (\ref{tss9}) and
(\ref{tss10}) into eq. (\ref{gen11}) yields simple kinetic equations
\begin{equation} \label{tss12}
\dot \rho_{\mu\mu'} = - w_{01}\rho_{\mu\mu'}, \;\;\mbox{with}\;\;
\mu\mu' = 01, 10.
\end{equation}
describing the exponential $0\!-\!1$ dephasing with the rate
\begin{equation} \label{tss13}
w_{01} = \mbox{$\frac{1}{2}$}w_{11} =
\mbox{$\frac{1}{2}$}J(\omega_s)(1+e^{-\beta\omega_s})^{-1}.
\end{equation}

The analogous consideration enables one to derived the kinetic
equations for relaxation of $0\!-\!2$ phases:
\begin{equation} \label{tss14}
\dot \rho_{\mu\mu'} = - w_{02}\rho_{\mu\mu'}, \;\;\mbox{with}\;\;
\mu\mu' = 02, 20,
\end{equation}
where
\begin{equation} \label{tss15}
w_{02} = \mbox{$\frac{1}{2}$}w_{22} =
\mbox{$\frac{1}{2}$}J(\omega_s)(1+e^{\beta\omega_s})^{-1}.
\end{equation}

It is important to note that in the case of irreversible $1 \to 2$
relaxation transitions, corresponding to the limit $\beta \omega_s
\to \infty$, the obtained kinetic equations can be represented in a
simple matrix form with the use of the operator $P_{1} = |1\rangle
\langle 1|$ of projection onto the reactive state $|1\rangle$:
\begin{equation} \label{tss15a}
\dot \rho = - \mbox{$\frac{1}{2}$} w_{11}^{\infty}[P_1,\rho]_{+},
\end{equation}
with $w_{11}^{\infty}$ defined in eq. (\ref{tss8}) and
$[P_1,\rho]_{+} = P_1 \rho + \rho P_1$. It is easily seen that eq.
(\ref{tss15a}) correctly describes relaxation not only of the
diagonal elements of the density matrix (population relaxation) but
also of all non-diagonal elements (phase relaxation).

Naturally, the effect of additional irreversible relaxation
transitions $0\to 2$ with the rate $w_{00}^{\infty}$, can also be
described, but with the more general equation $\dot \rho =
-\mbox{$\frac{1}{2}$}( w_{11}^{\infty}[P_1,\rho]_{+} +
w_{00}^{\infty}[P_0,\rho]_{+}),$ with $P_{0} = |0\rangle \langle
0|$. Of course this equation predicts additional dephasing in the
system of type of that considered above.

Note that similar expression, can be written for reversible
transitions (in the case of finite $\beta \omega_s$). One should
only add the terms proportional to $P_{2} = |2\rangle \langle 2|$
and the operator describing $1\!-\! 2$ (or $0\!-\! 2$) relaxation.

\subsection{Effect of splitting fluctuations}

In the above analysis we have not taken into account possible
fluctuating interactions diagonal in the bases $(|0\rangle,
|1\rangle, |2\rangle)$ (the diagonal contributions to  $H_{sl}$).

In the two state system considered above (see Sec. IIIA) the effect
of the fluctuating diagonal interaction of type of
$\mbox{$\frac{1}{2}$} \Phi_z ({\bf q}) \sigma_z$ is known to reduce
to additional dephasing.\cite{Blo,Hbr,Abr,Alek}

In particular, in the simple case of isotropic fluctuating
interaction $\{\Phi_n\}$, $(n = x,y,z)$, resulting in the
interaction $H_{sl} = \mbox{$\frac{1}{2}$} \!\sum_{n = x,y,z}
\!\Phi_{n}({\bf q})\sigma_{n}$, i.e. for $J_{nn'} (\omega) =
J(\omega) \delta_{nn'} , \;\; (n,n' = x,y,z)$, the  dephasing rate
$w_n = w_{12}$ is represented as \cite{Abr,Alek}
\begin{equation} \label{tss16}
w_{n}  = \mbox{$\frac{1}{2}$}({\bar w}_{n} + w_{11}),
\;\;\mbox{with}\;\; {\bar w}_{n} = J(0).
\end{equation}
Comparison of this formula with eq. (\ref{tss7}) shows that the
above-mentioned additional dephasing manifests itself in the
contribution $\mbox{$\frac{1}{2}$}w_{n}^{0} =
\mbox{$\frac{1}{2}$}J(0)$ to the dephasing rate.

Naturally, similar effect is predicted in the presence of the
additional fluctuating diagonal interaction, changing $(0-1)$
splitting:
\begin{equation} \label{tss17}
H_{sl}(01) = \mbox{$\frac{1}{2}$}(|0 \rangle \langle 0| - |1\rangle
\langle 1 |)\Phi_{01} ({\bf q}).
\end{equation}
Assuming uncorrelated fluctuations of $\Phi_{01}$ and the field
components $\Phi_{n}$ [see eq. (\ref{tss2})] one obtains for the
rate of $0\!\!-\!\!1$ dephasing
\begin{equation} \label{tss18}
w_{01} =  \mbox{$\frac{1}{2}$}({\bar w}_{01} + w_{11}),
\;\;\mbox{where}\;\; {\bar w}_{01} = J_{01}(0).
\end{equation}
The Fourier transformed correlation function $J_{01} (\omega)$ can
be calculated by substituting $\Phi_{01} ({\bf q})$ into eqs.
(\ref{gen2})-(\ref{gen9}).

Similar expression can be derived for the rate of $0\!\!-\!\!2$
dephasing caused by the additional fluctuating diagonal interaction,
changing $(0-2)$ splitting.

As in the absence splitting fluctuations, considered above, the
obtained relaxation equations are conveniently combined into matrix
equation which in the irreversible relaxation limit $\beta \omega_s
\to \infty$ is expressed in the form
\begin{equation} \label{tss19}
\dot \rho = - \mbox{$\frac{1}{2}$}\{w_{11}^{\infty}[P_1,\rho]_{+} +
{\bar w}_{01}\hat {\cal P}_{1} (\rho)\}.
\end{equation}
where
\begin{equation} \label{tss20}
\hat {\cal P}_{1}(\rho) = \mbox{$\frac{1}{2}$}[P_1,\rho]_{+} - P_1
\rho P_1.
\end{equation}
is the operator (in the Lindblat form\cite{Brag}) describing
$0\!-\!1$ and $1\!-\!2$ dephasing. This fact becomes especially
clear from the relation $\hat {\cal P}_{1}(\rho) = \frac{1}{2}(P_1
\rho Q_1 + Q_1 \rho P_1)$, where $Q_1 = 1 - P_1$.

The operator (\ref{tss20}) also predicts additional $1\!-\!2$
dephasing. This effect, however, is not relevant to the problem
under study and will not be discussed in this work.

\section{Radical pair recombination}

In this Section we will apply the results, obtained above, to the
analysis of the irreversible-reaction effect on the kinetics of RP
recombination. This process is known to be spin selective, i.e. the
reaction rate depends on the total electron spin ${\bf S} = {\bf
S}_a + {\bf S}_b$ of two radicals $a$ and $b$. The corresponding
states and transitions (as well as additional $ST$ dephasing with
the rate $\kappa_{ST}$) are schematically shown in Fig. 1b.

The spin/space evolution of the RP is described by the spin density
matrix $\rho$ in the space of states of the total electron spin: the
triplet states $|T_{+}\rangle, |T_{-}\rangle, |T_{0}\rangle$ and the
singlet state $|S\rangle$. In the simplest kinetic approach this
density matrix satisfies the Liouville equation \cite{St,Fre}
\begin{equation} \label{rpr0}
\dot \rho = -i [H_{p},\rho] - \hat K\rho,
\end{equation}
in which $ H_{p} = H_{a} + H_{b}$ is the spin Hamiltonian of the
pair of non-interacting radicals (its particular form is not
important for our further discussion) and $\hat K$ is the
reaction/relaxation supermatrix, i.e the matrix in the space of
matrix elements $|\nu\nu'\rangle \equiv |\nu\rangle \langle\nu'|, \;
(\nu,\nu' = S, T_{\mu} \;\mbox{with}\;\mu = \pm,0)$.

In the BRA the general expression for $\hat K$ can be found by
straightforward generalization of results, obtained in Sec. III [see
eqs. (\ref{tss19}) and (\ref{tss20}]. In what follows we will
consider the case of irreversible reaction in both $|S \rangle$ and
$| T_{\mu} \rangle$ states with rates $\kappa_S$ and $\kappa_T$,
respectively, and additional $ST$ dephasing with the rate
$\kappa_{ST}$, expected to be induced by the fluctuating exchange
interaction. The rate $k_{T}$ of reaction in $ |T_{\mu} \rangle$
states, as well as the rate $\kappa_{ST}$ of $ST_{\mu}$ dephasing,
are assumed to be independent of the spin projection, i.e. of $\mu,
$ ($\mu =0,\pm$).

Under these simplifying assumptions, in accordance with results of
Sec. III, the supermatrix $\hat K$ can, in general, be written as
\begin{eqnarray}
\hat K \rho \!&=&
\!\mbox{$\frac{1}{2}$}\big\{\kappa_{S}[P_S,\rho]_{+} +
\kappa_{T}[P_T,\rho]_{+} + \kappa_{ST}\hat {\cal P}_{S}(\rho)\big\}
\label{rpr1a}\\
&=& \!\mbox{$\frac{1}{2}$}\big\{\kappa_{T}\rho
+(\kappa_{S}-\kappa_{T})[P_S,\rho]_{+} + \kappa_{ST}\hat {\cal
P}_{S} (\rho)\big\}, \quad\;\; \label{rpr1b}
\end{eqnarray}
where
\begin{equation} \label{rpr2}
P_S = |S\rangle \langle S|, \;\;P_T = 1 - P_S = \!
\sum\nolimits_{\mu = \pm,0}\!|T_{\mu}\rangle \langle T_{\mu}|,
\end{equation}
are operators of projection  on $S$ and $T$ states, and
\begin{equation} \label{rpr3}
\hat {\cal P}_{S} (\rho) = \mbox{$\frac{1}{2}$}[P_S,\rho]_{+} - P_S
\rho P_S = \mbox{$\frac{1}{2}$}(P_S \rho P_T + P_T \rho P_S)
\end{equation}
is the term describing additional $ST$ dephasing. Note that $ \hat
{\cal P}_{S} (\rho)= \hat {\cal P}_{T} (\rho) =
\mbox{$\frac{1}{2}$}[P_T,\rho]_{+} - P_T \rho P_T $.

In accordance with general analysis in Sec. III and formula
(\ref{rpr1a}) the reaction/relaxation rates
\begin{equation} \label{rpr4}
k_{\nu\nu'} = \langle \nu\nu' |\hat K| \nu\nu' \rangle,\;
\;(\nu,\nu' = S, T_{0,\pm}):
\end{equation}
are written as
\begin{eqnarray}
k_{SS} &=&  \kappa_{S}, \; \; k_{T_{\mu}T_{\mu}} = \kappa_{T}, \:
(\mu = 0,\pm), \qquad \label{rpr5a}\\
 k_{ST_{\mu}} &=&  k_{T_{\mu}S} = \mbox{$\frac{1}{2}$}(\kappa_S +
\kappa_T + \kappa_{ST}). \quad \label{rpr5b}
\end{eqnarray}
Note that $k_{T_{\mu}T_{\mu}}$ and $k_{ST_{\mu}}$ are independent of
$\mu$.

In our derivation of these expressions we did not specify the final
states for reactive transitions from $S$ and $T$ states. These final
states are, quite probably, different for the initial $S$ and $T$
states. However, the choice of final states does not, evidently,
affect the results in the considered case of irreversible reactions.

The rates $\kappa_S, \,\kappa_T$, and $\kappa_{ST}$ can, in
principle, be evaluated by means of the expressions, presented in
Secs. II and III, in terms of correlation functions of corresponding
interactions. Unfortunately, for many systems, of type of those
discussed in our work, the obtained general expressions appear to be
not very useful for accurate calculations of the rates because of
the complexity of the systems which results, in particular, in the
complexity of evaluating correlation functions $K_{nn'}^{\pm}$.
Nevertheless, the expressions are quite helpful for qualitative and
semiquantitative analysis, as it will be demonstrated below.

It is easily seen that in the absence of additional $ST$ dephasing
($\kappa_{ST} = 0$) equation (\ref{rpr1a}) reduces to the
conventional formula (\ref{intr1}), thus showing that this formula
is quite rigorous and is valid within the region of applicability of
the BRA.

As for the applicability of the BRA as applied to the RP
recombination, according to the general comments presented in the
end of Sec. II the BRA is valid when $\|\hat K \|\tau_c \ll 1$,
where $\tau_c$ is the characteristic correlation time of
fluctuations of $H_{sl}$. The time $\tau_c$ can roughly estimated as
the inverse characteristic frequency of vibrational and/or
rotational motions which determine $H_{sl}$ fluctuations. The most
reasonable estimation for this time is $\tau_c \sim 10^{-13}
s^{-1}$. Taking into account that usually $\kappa_S, \kappa_T,
\kappa_{ST} \lesssim 10^{11} s^{-1}$, we arrive at the estimation
$\|\hat K \|\tau_c \lesssim 10^{-2}$ which ensures good accuracy of
the BRA. In fact, this inequality justifies quite natural
requirement that the probability of relaxation transitions to the
ground state during the characteristic time $\tau_c$ should be small
enough.

Of certain interest is the special property of the model
(\ref{intr1}), in which the density matrix $\rho (t)$ [obeying eq.
(\ref{rpr0})] is represented in the form $\rho (t) = |\psi
(t)\rangle \langle \psi (t) |$, with the "wave" functions $|\psi
(t)\rangle$ and $\langle \psi (t) |$ satisfying the
Shr\"odinger-like equations $d|\psi\rangle/dt = -[iH_p +
\mbox{$\frac{1}{2}$} (\kappa_S P_S + \kappa_T P_T)] |\psi\rangle$
and $d\langle \psi |/dt = -\langle \psi|[-iH_p +
\mbox{$\frac{1}{2}$} (\kappa_S P_S + \kappa_T P_T)] $.

It is worth noting that the limit of negligibly weak additional $ST$
dephasing, when $\kappa_{ST} \ll \kappa_S, \kappa_T$, can hardly be
considered as quite realistic, because of expected small values of
reaction rates in the case of large splitting $\omega_s$ of excited
(reactive) $S, T$ states and the ground state, which is typical for
the considered process. The fact is that, according to eqs.
(\ref{tss19}) and (\ref{tss20}), the reactive transition rates
$\kappa_{\nu}, \, (\nu = S, T),$ are determined by the Fourier
transforms $J_{\nu}(\omega_s)$ of the correlation function of the
interaction, inducing corresponding transitions: $\kappa_{\nu} \sim
J_{\nu}(\omega_s)$. The functions $J_{\nu}(\omega_s)$ rapidly
decrease when increasing $\omega_s \tau_c$, where $\tau_c$ is the
correlation time of interaction fluctuations. Typically $\omega_s $
is of order of the splitting of electronic terms, while $\tau_c$
(for vibronic coupling) is expected to be of order of (or larger
than) inverse vibrational frequencies $\omega_v \ll \omega_s$, so
that $\omega_s \tau_c \gg 1$ and therefore we get for the total $ST$
dephasing rate $\kappa_{ST}$ [see eq. (\ref{rpr5b})] the relation
$\kappa_{ST} > \kappa_S, \kappa_T$ and therefore $k_{ST_{\mu}} >
k_{SS}, k_{T_{\mu}T_{\mu}}$.

To illustrate these general arguments, note that typically the rate
$ \kappa_S \sim 10^{10} \div 10^{11} s^{-1}$ and $\kappa_T <
\kappa_S$. As for the dephasing rate $\kappa_{ST}$, it can be
estimated assuming by the simplest BRA
relation\cite{Blo,Red,Hbr,Abr} $\kappa_{ST} \sim {\bar J}^2 \tau_c$,
where ${\bar J}$ is the amplitude of fluctuations of the exchange
interaction due to relative motion of radicals in the well of the
attractive interradical interaction potential and $\tau_c \sim
10^{-13} s$. Assuming that $\bar J \sim J_0 (\alpha\lambda)$, where
$J_0 \sim 10^{14} s^{-1}$ is the exchange interaction between
radicals at a contact interradical distance $r = d$, $\alpha \sim
10^8\, {\rm cm}^{-1}$ is the rate of decrease of this interaction
with the distance $r$: $J_e(r) = J_0 e^{-\alpha (r-d)}$, and
$\lambda \gtrsim 10^{-10}\, {\rm cm}$ is the estimated amplitude of
stochastic relative motion of radicals at short distances.
Substitution of all these parameters into formula for $\kappa_{ST}$
yields $\kappa_{ST} \gtrsim 10^{11}\, s^{-1} \:(\gtrsim \kappa_{S},
\kappa_{T})$.

The additional arguments in favor of fast $ST$ dephasing,
corresponding to the relation $k_{ST_{\mu}} > k_{SS},
k_{T_{\mu}T_{\mu}}$, can be obtained by the analysis of the effect
of stochastic (diffusive) relative motion of radicals with the use
of the stochastic Liouville equation.\cite{Shu2,Shu3} The fact is
that so far in our analysis we have not specified the mechanism of
stochastic motion along lattice ${\bf q}$-coordinates which lead to
fluctuations of $H_{sl}$, though in our above estimations we have
implied that the coordinates ${\bf q}$ correspond to the localized
rotational/vibrational motion of the RP complex (cage) at the short
distances $r \sim d$. In reality, however, very often one should
take into account the additional coordinates describing relative
diffusion of particles in the area of large space. In many case the
contribution of these coordinates to the relaxation process in the
RP  appears to be more important than that of rotational/vibrational
motion at short distances.\cite{Shu2,Shu3}

Naturally, diffusion leads to the stochastic change of the
interradical distance $r$ and therefore to fluctuations of distance
dependent interactions and kinetic parameters. In describing the
manifestation of relative diffusive motion the effect of reaction
and relaxation at short distances can be modeled by
reaction/relaxation term $\hat K (r)\rho$ similar to that in eq.
(\ref{rpr0}) but with the distance dependent supermatrix $\hat K
(r)$, i.e. with the supermatrix (\ref{rpr1a}) in which the rates
$\kappa_{\nu} (r), \: (\nu = S, T, ST)$ are distance dependent. This
model is expected to be fairly accurate since the characteristic
scale of these $r$-dependences is $\alpha^{-1} \sim 10^{-8}\, {\rm
cm}$, while the typical jump length $l_D$ of the diffusive motion of
not very small molecules is smaller than $\alpha^{-1}$: $l_D
\lesssim 10^{-9} \,{\rm cm}$. The diffusive relative motion also
leads to fluctuations of the exchange interaction $J_{ex}(r)$ at
distances larger than the contact distance $d$ (the distance of
closest approach). The fluctuations of $J_{ex}(r)$ result, in turn,
in those of $ST$ splitting and therefore in $ST$ dephasing.

The analysis with the stochastic Liouville equation
shows\cite{Shu2,Shu3} that the efficiency of (distance dependent)
reactivity and exchange interaction is characterized by the
corresponding radii. In particular, the reaction radii $l_{\nu\nu}\,
(\nu = S,T),$ for $S$ and $T$ states resulting from reactivities
$\kappa_{\nu} (r) = \kappa_{\nu}^0 \lambda_0 \delta (r-d), \: (\nu =
S,T)$, localized in the narrow region $\sim\lambda_0$ near the
distance of closest approach $d$, can be evaluated by\cite{Shu1}
\begin{equation} \label{rpr6}
l_{\nu\nu} =d [q_{\nu}/(1 + q_{\nu})],\:(\nu = S,T),
\end{equation}
where  $q_{\nu} = \kappa_{\nu}^0 (d\lambda_0/D)$ and $D$ is the
coefficient of relative diffusion of radicals. As for the dephasing
radius, it can be evaluated assuming exponential distance dependence
$J_e (r) = J_0 e^{-\alpha (r-d)}$. In the most realistic limit of
relatively strong interaction, when $J_0 > D\alpha^2$, the $ST$
dephasing radius $l_{ST}$ is independent of dephasing at a contact
and is given by\cite{Shu1}
\begin{equation} \label{rpr7}
l_{ST} = l_{TS} \approx d + \alpha^{-1}[1.14 +\ln
(2|J_0|/D\alpha^2)].
\end{equation}
i.e. for $|J_0| > D\alpha^2$ the dephasing radius $l_{ST} > l_{SS},
l_{TT}$.

Noteworthy is that since the rates of reaction in the states
$T_{\mu}, \, (\mu = 0, \pm)$ and dephasing rates, predicted by the
supermatrix (\ref{rpr1a}), are independent of the projection $\mu$
and therefore the reaction radii $l_{T_{\mu}T_{\mu}}$ and
$l_{ST_{\mu}} = l_{T_{\mu}S}$ are also independent of $\mu$:
$l_{T_{\mu}T_{\mu}} \equiv l_{TT}$ and $l_{ST_{\mu}} \equiv l_{ST}$.

Most clearly the meaning of this relation between radii can be
demonstrated as applied to diffusion assisted processes in cages:
potential wells, micelles, etc. The kinetics of these processes is
known to be close to exponential\cite{Shu2,Shu3,Shu4} and the time
evolution of spin density matrix is described by the kinetic
equation of type of eq. (\ref{rpr0}) with (diagonal) supermatrix
$\hat K$ whose elements $k_{\nu\nu'}$ can be estimated in terms of
corresponding radii\cite{Shu2,Shu3}
\begin{equation} \label{rpr8}
\langle \nu\nu' |\hat K| \nu\nu'\rangle = k_{\nu\nu'} \approx
Dl_{\nu\nu'}/Z, \;\:(\nu = S, T_{0,\pm}),
\end{equation}
where $Z = \int_{r \in c} dr\, r^2 e^{-\beta u(r)}$ is the partition
function of particles within the cage [or in the potential well
$u(r)$]. Recall that the radii $l_{T_{\mu}T_{\mu}},\, (\mu =
0,\pm),$ and $l_{ST_{\mu}}$  are in dependent of $\mu$ so that
$k_{T_{\mu}T_{\mu}}$ and $k_{ST_{\mu}}$ do not depend on $\mu$
either: $k_{T_{\mu}T_{\mu}} \equiv k_{TT}, \: k_{ST_{\mu}} =
k_{T_{\mu}S} \equiv k_{ST}$.

Taking into account the above-mentioned relation between the radii,
we get $k_{ST} > k_{SS}, k_{TT}$. In terms of the rates of the THSS
relaxation supermatrix (\ref{rpr1a}), (\ref{rpr1b}) this inequality
means that $\kappa_{ST}
> \kappa_{S}, \kappa_{T}$. In reality, the dephasing rate $k_{ST}$
can essentially larger than $k_{SS}$ and $k_{TT}$ since the
corresponding radius $l_{ST}$ increases with the decrease of the
diffusion coefficient $D$ (though this dependence is very slow). For
example, for realistic values\cite{Fre} $J_0 = 10^{12} \, {\rm
s}^{-1}$ (low but still quite realistic estimation), $\alpha =
10^{8}\, {\rm cm}^{-1}$, and $D = 10^{-5}\, {\rm cm}^2/{\rm s}$ one
gets $l_{ST} - d \approx 4.14 \cdot 10^{-8}\, {\rm cm}$. Taking into
account that usually $d \approx 3\div 5 \cdot 10^{-8}\, {\rm cm}$ we
conclude that the rate of dephasing can be about twice as large as
that of reaction.

It is important to emphasize that in the realistic strong exchange
interaction limit  $|J_0| > D\alpha^2$ the radius $l_{ST}$ and,
therefore, the rate $k_{ST}$ is independent of $\kappa_{ST}$. This
means that in the strong interaction limit the above discussion of
specific features of the superoperator $\hat K$ and, in particular,
the relation between reaction and relaxation rates is not quite
relevant. The fact is that in this limit the effect of dephsing is
determined by fluctuations of the exchange interaction at large
distances $r \sim l_{ST}$ governed by relative diffusion of
radicals.

Note that different possible models of the recombination process,
including the models based on quantum measurement formalism,
correspond to different relations between $k_{ST_{\mu}} $, $k_{SS}$,
and $k_{T_{\mu}T_{\mu}}$ and therefore can interpreted within BRA
type approaches under certain assumptions on values of interactions
in the system.

In the end of this Section it is also worth noting that the above
discussion of the form of the reaction/relaxation supermatrix $\hat
K$ is, of course, important for principle understanding the specific
features of the relaxation kinetics of spin selective processes. As
far as applications are concerned, however, the theoretical analysis
shows that for majority of processes in a wide region parameters the
particular relation between elements of this supermatrix only weakly
manifests itself in experimentally measured observables. For
example, for the probability of the spin dependent and diffusion
assisted RP recombination the effect of the specific relation
between $l_{SS}$ and $l_{ST}$ is characterized by the parameter $\xi
= Q (l_{ST}-l_{SS})^2/D$,\cite{Shu2,Shu3,Shu4} where $Q$ is the
characteristic spin dependent interaction in free radicals
(hyperfine interaction, $\Delta g$ change of the Zeeman frequences)
The small value of this parameter means that the probability is
insensitive to the particular relation between $l_{SS}$ and
$l_{ST}$. In particular, for realistic values $D = 10^{-5}\, {\rm
cm}^2/{\rm s}$, $l_{ST}-l_{SS} = 3\cdot 10^{-8}\,{\rm cm}$, and $Q =
10^{9} s^{-1}$ we get fairly small $\xi = 0.09$, which means that at
low viscosities (for highly mobile particles) the the effect of some
difference between $l_{SS}$ and $l_{ST}$ can hardly be observed
experimentally. As the viscosity increases, however, this effect
becomes quite measurable.

\section{Discussion and conclusions}

The main goal of this short paper is to demonstrate that
manifestations of state-selective reactivity in the kinetics of
relaxation of quantum systems can quite conveniently and physically
reasonably be treated within the BRA. It is important to note that
the BRA is a mathematically rigorous approach and absolutely
transparent from physical point of view. The BRA allows one to
analyze the contributions of different mechanisms and interactions
to the kinetics of the processes giving deep insight into its
specific features.

The important example of such processes is spin-selective reactions
of paramagnetic particles in which the reaction kinetics is
essentially governed by (quantum) evolution of the spin subsystem
via spin-dependent reactivity.

In our works we have illustrated the obtained general results by the
example of RP recombination.

In the simplest assumptions on the mechanism of the spin-selective
reactivity the BRA is shown to predict the conventional expression
(\ref{intr1}) for the reactivity supermatrix $\hat K$. Its physical
meaning can quite clearly and rigorously be demonstrated within the
simple variant of the model (\ref{tss1}), (\ref{tss2}) with the
Hamiltonian  $H_s$ given by eq. (\ref{tss1}) but $H_{sl}= v(t)
(|1\rangle \langle 2| + |2\rangle \langle 2|) $, where $v(t)$ is a
real fluctuating function with zeroth mean: $\langle  v (t)\rangle =
0$. Solution of the corresponding Schr\"edinger equation for the
wave function $|\psi(t)\rangle = \sum_{j = 0,1,2} a_{j}(t)
|j\rangle$ in the second order in $v(t) t \ll 1$ yields for the
initial condition $a_{1} (0) = a_{0} (0) = 1, \, a_{2} (0) = 0$:
$a_{1} (t) \approx e^{-i\omega_s t/2} [1 - \Delta a (t)],$ where $
\Delta a (t) = \int_0^{t}dt_1 \int_0^{t_1}dt_2 v(t_1)v(t_2)
e^{i\omega_s (t_1 -t_2)}$, and $a_0 (t) = e^{-i\omega_0 t}$. Thus
one gets the following expressions for the density matrix elements
averaged over $v(t)$-fluctuations:\cite{Shu5} $\rho_{11} (t) =
\langle |a_{1} (t)|^{2} \rangle \approx  1 - 2{\rm Re} \langle
\Delta a(t) \rangle$ and $\rho_{01} (t) = \langle a_{0}^{*} (t)
a_{1} (t) \rangle \approx [1 - \langle \Delta a(t)
\rangle]e^{i(\omega_0 -\omega_s/2)t}$. Note that for relatively
small $t$, satisfying the relation $\tau_c < t \ll 1/\sqrt{\langle
v^{2}\rangle}$, where $\tau_c$ is the correlation time of $v
(t)$-fluctuations, we can write the relation $2{\rm Re}\Delta a(t)
\approx w_{11} t$ in which $w_{11} = 2\int_0^{\infty}dt \langle
v(t)v(0) \rangle \cos (\omega_s t) $ is $(1 \to 2)$-transition rate.
The above formulas show also the rate of dephasing $w_{01}$, i.e.
the rate of decay of $\rho_{01} (t)$, is given by $w_{01}
\stackrel{t\to\infty}{=} {\rm Re}\Delta a(t)/t = \frac{1}{2}w_{11}$,
in agreement with formula (\ref{intr1}).

The analysis of the obtained simple expressions demonstrates that
the relation $w_{01} = \frac{1}{2}w_{11}$ [i.e. eq. (\ref{intr1})]
actually results from the evident fact that, if $\rho_{11} (t) \sim
|a_{1} (t)|^2 \sim e^{-w_{11}t}$, then $\rho_{01} (t) \sim a_{1} (t)
\sim e^{-w_{11}t/2}$.

The BRA analysis of general mechanisms, taking into account possible
strong dephasing which can accompany the reaction process, has
enabled us to essentially generalize formula (\ref{intr1}). The
generalized BRA permits natural physical interpretation of all
results, obtained within recent "advanced"
approaches,\cite{Hor,Kom,Il} in terms of clear physical parameters
and under reasonable assumptions.

For example, recently proposed mechanism, based on some ideas of the
quantum measurement theory, predicts the $ST$ dephasing with the
rate twice as large as that following from eq. (\ref{intr1}), i.e.
with the rate $k_{ST}^0 = k_{SS} + k_{TT}$ rather than with
$k_{ST}^0 = \frac{1}{2}(k_{SS} + k_{TT})$.\cite{Hor} This means, in
particular, that in the case of only one reactive state, say singlet
($S$), $ST$ phase is predicted to decay with the reaction rate:
$k_{ST} = k_{SS}$. Attempting to apply this result to real
processes, one should keep in mind that, first, the applicability of
the measurement theory\cite{Brag} to the processes at molecular
scale is not quite evident\cite{Home} and, second, any value of the
$ST$ dephasing rate can easily be interpreted in much more clear way
by the analysis of fluctuating intermolecular interactions within
the BRA. Note also that just the relation $k_{ST} = k_{SS}$ is
predicted by the most realistic diffusion theory, briefly discussed
above. The fact is that according to eqs (\ref{rpr6})-(\ref{rpr8})
in the limits of high reactivity  $k_{S}^0 (d\lambda/D) \gg 1, $ and
not very strong exchange interaction $J_0/(D\alpha^2) \sim 1$ one
gets $k_{SS} \equiv k_{S}\approx k_{ST} \approx d$ (in this
estimation we also took into account that usually $\alpha d \gg 1$).

Similar comments can be added to the results of recent
works\cite{Kom,Il} concerning the study of RP recombination within
the models, which can be considered as some variants of the spin
boson model,\cite{Talk} usually applied in the theory of interaction
of quantum dots. The model used in ref. [5] predicts that RP
interaction results in the reactivity term of the form $\hat K \rho
\sim \hat {\cal P}_{S} (\rho) \sim [P_S,\rho]_{+} - 2 P_S \rho P_S$
which, as it follows from eq. (\ref{rpr3}), implies $ST$ dephasing
without reaction (recombination). This result means that in the
model, used in ref. [5], the process reduces to $ST$ splitting
fluctuations, naturally, resulting only in $ST$ dephasing. The
author treats this dephasing as a set of measurement events which,
in accordance with the Von Neumann "reduction postulate" (or
"collapse postulate"),\cite{Home,Fac} are associated with the
instant loss of phase. This interpretation, however, does not look
quite convincing. Another spin-boson type model (with another
Hamiltonian) is considered in the work [19]. This model predicts for
$\hat K$ formula (\ref{intr1}) which means (in terms of our general
analysis) that in the corresponding fluctuating part of the
Hamiltonian the diagonal part, describing the fluctuations of
splitting of terms, is neglected. In addition to the mentioned
limitations of the analysis in works [5,6], note that as applied to
the RP recombination any spin boson type models, including those
applied in refs. [5,19] are of fairly restricted applicability
(mentioned by the authors), for example, in the case of strongly
unharmonic or stochastic (i.e diffusion like) relative motion of
particles. Note also that this kind of models is too complicated to
be useful for the analysis of contributions of different kinds of
interactions as well as the simultaneous effects of localized and
delocalized motions.

Moreover our analysis of the RP recombination within the BRA and
diffusion approximation shows that this process can easily and
rigorously be described by direct studying the kinetics of quantum
relaxation without attracting any additional ideas whose
applicability to small systems of molecular scale has not be
rigorously demonstrated so far, and (what is most important) which,
in fact, do not provide any new insight into the problem in addition
to that obtained above from rigorously derived kinetic equations.

Concluding this short discussion we would like emphasize once more
that the effect of the state-selective reactive decay on relaxation
kinetics of quantum system can fairly easily and accurately be
estimated within the BRA. This estimation is usually sufficient for
majority of interpretations. Of course, in some cases more detailed
analysis within more specific models is required, but the models
based on general ideas of the quantum measurement theory can hardly
help in this kind of analysis.

\textbf{Acknowledgements.}\, The author is grateful to Dr. V. P.
Sakun for valuable discussions. The work was partially supported by
the Russian Foundation for Basic Research.





\end{document}